# Multicore fiber delay line performance against bending and twisting effects

Sergi García, Mario Ureña, Rubén Guillem and Ivana Gasulla
ITEAM Research Institute, Universitat Politècnica de València, Spain, sergarc3@iteam.upv.es

**Abstract** *We report the experimental evaluation of bending and twisting effects on the propagation and radiofrequency signal processing performance of multicore fibers. We demonstrate that twisting minimizes to a large extend the group delay fluctuations over a bent multicore fiber.*

## Introduction

One of the major detrimental effects on weakly-coupled multicore fiber (MCF) transmission is related to possible bending and twisting effects that may affect deployed fiber links. Although these effects have been studied mainly in relation to the fiber intercore crosstalk[1,2] (both in homogeneous and heterogeneous MCFs), a thorough analysis on the group delay performance of the different cores is required for certain applications where time-delay control and synchronization play a crucial role. This is the case, for instance, of MCF-based radio access networks feeding multiple input multiple output (MIMO) antenna systems in the upcoming multigigabit-per-second 5G wireless communications[3]. Another representative scenario is Microwave Photonics signal processing, where optical architectures based on discrete-time approaches are assembled to generate or modify radiofrequency (RF) signals[4]. Here, MCFs have been proposed as a compact medium to implement different processing functionalities[5] such as reconfigurable signal filtering, optical beamforming for phased-array antennas or optoelectronic oscillation[6].

In this paper, we evaluate the effects of fiber bending and twisting on the group delay characteristics of a homogeneous MCF link and how they reflect on the performance of delay-sensitive signal processing applications. We experimentally demonstrate, for the first time to our knowledge, how twisting counteracts the group delay alterations introduced by curvatures.

## Evaluation of bending and twisting effects on the MCF group delay performance

A bent core $m$ can be described as a corresponding straight core fiber with an equivalent refractive index distribution[7] $n_{eq,m}$:

$$n_{eq,m}^2 = n_m^2\left(1 + 2\frac{r_m}{R_b}\cos\theta_m\right), \qquad (1)$$

where $R_b$ is the bending radius, $n_m$ is the refractive index of core $m$ in the straight fiber and $(r_m,\theta_m)$ are the local polar coordinates of core $m$, (see Fig. 1). From Eq. (1), we see that the highest variation on the refractive index (and thus the group delay) due to curvatures may occur when the core forms an angle with the curvature plane of $\theta = k\pi$, $k \in Z$. Eq. (1) also shows that cores located in the 1st and 4th quadrants of the fiber cross section show an opposite behavior than those located in the 2nd and 3rd ones.

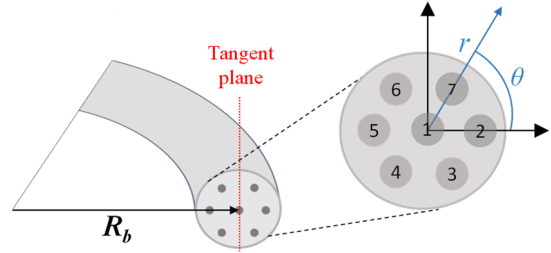

**Fig. 1:** Fiber curvature with a bending radius $R_b$ and local polar coordinates $(r,\theta)$ indicated in the MCF cross section.

Fiber twisting with a constant twist rate $\gamma$ can be understood as a linear rotation of the fiber cross section along the fiber length $L$. This can be interpreted as a linear increment of the angle $\theta_m$ for each core $m$. If we apply an ideal curvature with a fixed bending radius over the whole fiber length, the angle $\theta_m$ of each core would be preserved. The differential group delay (DGD) between the outer and the central cores would then accumulate linearly with the fiber length unless the core is placed at an angle $\theta = \pi/2 + k\pi$, $k \in Z$.

We have experimentally evaluated both bending and twisting effects on the group delay performance of a commercial homogeneous 7-core fiber. The cladding diameter is 125 μm and the core pitch is 35 μm. Different bending conditions were investigated in comparison to the straight condition. To properly maintain the MCF straight and avoid unintentional twists, we restricted the fiber length to 3 meters. Small bending radii of 35 and 75 mm are then used to force a representative variation in the core group delays of the bent fiber.

We have measured the DGD between each outer core and the central core by using an interferometric-based technique[8]. Variable delay lines at the output of each core are used to set the DGDs between cores to zero when the fiber is straight and thus suppress the residual differences due to undesired effects such as fabrication mismatches. Figure 2 gathers the measured DGDs between the outer and the central cores for the following scenarios: (a) a 35-mm bend without any intentional twist, (b) a 35-mm bend with a forced twist rate of approximately $\gamma = 2\pi$/m (i.e., a total twist of $\varphi_t$ = 3 turns for the 3-meter fiber), and (c) a 75-mm bend without any intentional twist. Dashed lines represent the estimated transverse planes that are tangent to the curvature. We observe a slightly asymmetric behaviour between the cores with positive and negative delay variation, which can be attributed to mismatches on the core pitch (or core location) due to fabrication errors. As we have previously mentioned, the worst-case variation occurs when the core forms an angle of $\pm\pi/2$ with the tangent plane. From Fig. 2, we see how the absolute value of the DGD increases as the core is closer to this value, as expected from Eq. (1).

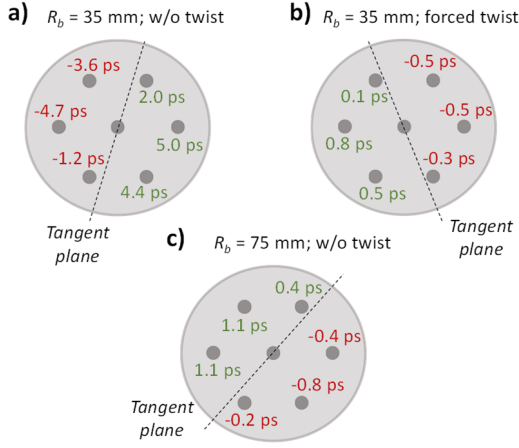

**Fig. 2:** Differential group delays measured between the central and outer cores for (a) a bending radius of $R_b$ = 35 mm without twist; (b) $R_b$ = 35 mm with a forced twist; and (c) $R_b$ = 75 mm without twist.

From Fig. 2 we can observe that the maximum DGD between cores decreases from 5 down to 1.1 ps as the bending radius increases, and decreases down to 0.8 ps when the twist is applied. This behaviour is illustrated more clearly in Fig. 3, where the dependence of the DGD with the bending radius and the total twist is shown. Fig. 3(a) shows the computed responses (lines) of the DGD as a function of the bending radius when different twist conditions are applied (total twist from 0 to 6.50$\pi$). The measured worst-case DGDs for $R_b$ = 35 mm with and without twist and $R_b$ = 75 mm without twist are depicted as markers. Although we take care to not cause any twist for the non-twist measurements, we see here that a residual twist of around 0.75$\pi$ seems to be applied. This can be caused by multiple factors, but we believe that the main contribution is fiber splicing onto the fan-in/fan-out devices. This residual twist causes an important reduction of around 70% and 80%, respectively, for the 35-mm and 75-mm bends. Figure 3(b) shows the dependence of the worst-case differential delay with the total twist along the 3-meter fiber. The computed behaviour for the 35-mm and the 75-mm bends are depicted in blue and orange lines, respectively. We see how both curves decrease rapidly as the twist is increased following a sinc shape. This behaviour can be derived from Eq. (1) if we express the maximum (or worst-case) DGD accumulated along the link length as:

$$\left|DGD_{accumulated}^{Worst\ Case}\right| = \frac{\tau_m r_m L}{R_b}\text{sinc}(\gamma L), \quad (2)$$

where $\tau_m$ is the group delay of core $m$ in straight condition. We can observe in Fig. 3b how the measured results fit well with their corresponding curves. When no twist was applied, we see that both measurements remain in the light-grey zone, which corresponds to a total twist below 1 turn (i.e., below 2$\pi$). On the other hand, the forced twist of 3 turns along the 3-meter fiber is well-matched with the computed curve. We can confirm then that the experimental results are in a good agreement with the expected values.

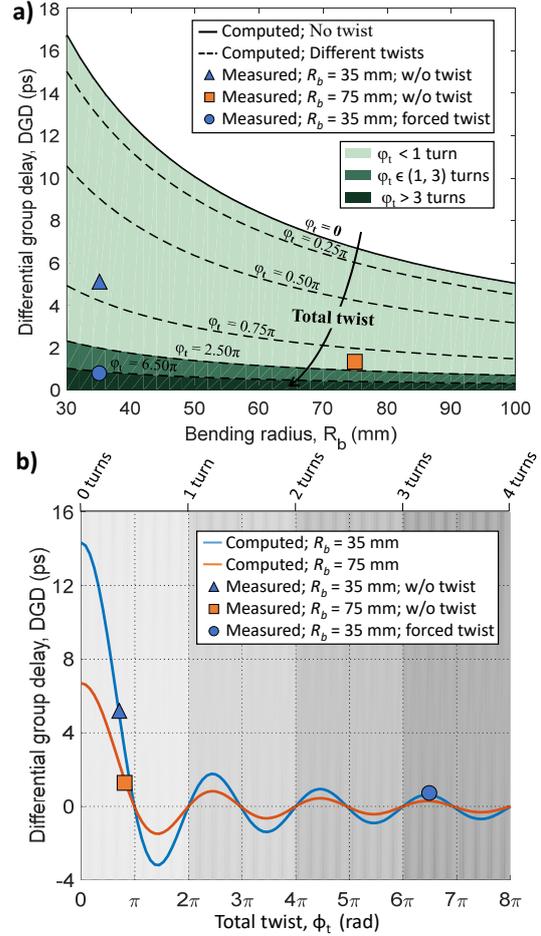

**Fig. 3:** Differential group delay dependence on (a) the bending radius and (b) the total twist. Solid and dashed lines correspond to the worst-case computed responses, while filled markers represent the worst-case experimental values.

**Evaluation of bending and twisting effects on the signal processing performance**

In particular, variations on the group delay difference between cores may affect applications where the MCF serves as a signal delay line. This is the case of RF signal processing, where the MCF offers the required parallelism for the implementation of a compact true time delay line architecture using a single optical fiber[5]. Therefore, we have experimentally evaluated how bending and twisting affect a delay-sensitive application such as Microwave Photonics signal filtering[4]. We implemented a 7-tap finite impulse response filter that is synthesized to have a Free Spectral range of 10 GHz by adjusting the group delay difference between adjacent filter samples (or basic differential delay) to $\Delta\tau$ = 100 ps. The filter frequency response was measured for four different fiber conditions: (1) straight; (2) bent with $R_b$ = 35 mm; (3) bent with $R_b$ = 75 mm; (4) bent with $R_b$ = 35 mm and twisted.

The RF filter setup is depicted in Fig. 4a. The optical signal is modulated by an electro-optic modulator (EOM) and equally launched to all the fiber cores using a 1:8 coupler. The signal is injected/extracted from/to the MCF using fan-in/fan-out devices. The required filter basic differential delay $\Delta\tau$ was achieved by adding 2 cm of singlemode fiber incrementally from tap to tap. We used a series of variable delay lines (VDLs) at the output of the MCF for the fine tuning of the tap delays. To obtain a uniform amplitude distribution between the filter taps, we use Variable Optical Attenuators (VOAs). After coupling together all the filter samples, we amplified the signal using an Erbium-doped fiber amplifier (EDFA) before photodetection.

Figure 4b illustrates the filter transfer function measured by an Electrical Spectrum Analyzer. The blue solid line represents the case with no bend, where the filter response shows clearly no degradation at all. The red dash-dotted line corresponds to the smallest bending radius ($R_b$ = 35 mm) and it shows how the fluctuations in the group delay affect visibly the filter response by increasing the sidelobe level. By contrast, the green dotted line displays how this degradation effect is compensated when an intentional twist is applied to the fiber. Finally, the yellow dashed line shows how a typical spool bending radius ($R_b$ = 75 mm) has practically no impact on the filter response, as expected from the group delay performance carried out in the previous section.

**Conclusions**

We have reported the experimental evaluation of intercore differential group delay variations induced by fiber bending and twisting that might affect MCF transmission. We demonstrated how twisting the fiber counteracts the group delay variations produced by curvatures even for short distances. These results will not only benefit high-capacity long-haul digital communications (where digital MIMO processing may be required to compensate intercore crosstalk), but also delay-sensitive application areas as fiber-wireless access networks and RF signal processing.

**Acknowledgements**

This research was supported by the ERC Consolidator Grant 724663, Spanish MINECO Project TEC2016-80150-R, Spanish scholarship MINECO BES-2015-073359 for S. García and Spanish MINECO Ramon y Cajal fellowship RYC-2014-16247 for I. Gasulla.

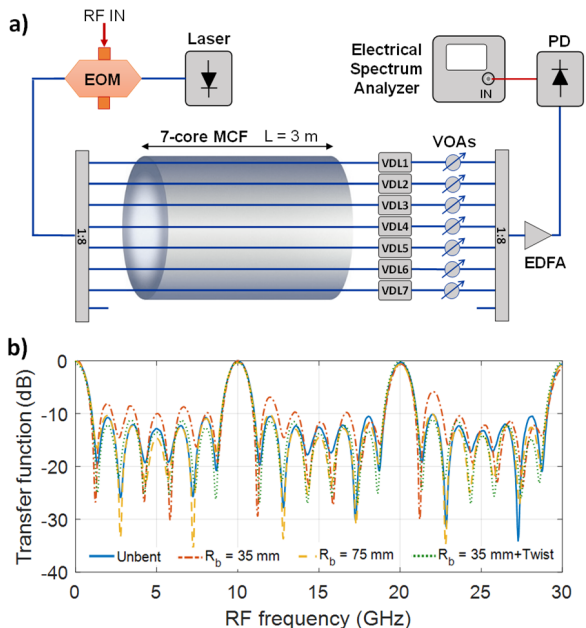

**Fig. 4:** (a) Experimental setup for RF signal filtering. (b) Measured filter transfer function for: (solid blue) unbent fiber, (dash-dotted red) bent with 35-mm radius, (dashed yellow) bent with 75-mm radius and (dotted green) bent with 35-mm radius and forced twist.